\documentclass[superscriptaddress, 12pt, aps, prl,  preprint, longbibliography]{revtex4-1}
\usepackage{graphicx}
\usepackage{epstopdf}
\usepackage{color}
\usepackage{amsmath, amssymb, bm}
\usepackage{physics}
\usepackage{hyperref}

\begin{document}
\newcommand{\lsh}[1]{[{\bf  \color{red}{LSH: #1}}]}

\title{Learning free energy across a differentiable temperature range by deep generative modeling}
\author{Yaowen Zhang}
\affiliation{Department of Physics,  Hong Kong University of Science and Technology, Hong Kong, China}
\author{Chen Chen}
\affiliation{Department of Physics,  Hong Kong University of Science and Technology, Hong Kong, China}
\author{Wanyu Lin}
\affiliation{Department of Computing, The Hong Kong Polytechnic University, Hong Kong, China}
\author{Shuo-Hui Li}
\email{shuohuili@ust.hk}
\affiliation{Department of Physics,  Hong Kong University of Science and Technology, Hong Kong, China}
\author{Ding Pan}
\email{dingpan@ust.hk}
\affiliation{Department of Physics,  Hong Kong University of Science and Technology, Hong Kong, China}
\affiliation{Department of Chemistry,  Hong Kong University of Science and Technology, Hong Kong, China}
\affiliation{IAS Center for AI for Scientific Discoveries, Hong Kong University of Science and Technology, Hong Kong, China}

\date{\today}

\begin{abstract}

Free energy governs the directionality, stability, and temperature sensitivity of physical, chemical, and biological transformations. Conventional free energy calculations, such as thermodynamic integration and free energy perturbation, often have to be performed at individual temperatures and are limited by the high cost of sampling an extensive set of thermodynamic states. Here, we introduce learned temperature-differentiable free energy perturbation (TD-FEP), a deep generative framework that learns a temperature-conditioned, invertible mapping between reference and target systems through data-free, energy-based optimization. TD-FEP thereby unifies one-shot state generation and free energy calculation across a continuous temperature range. We validated TD-FEP across a broad range of applications, including isolated molecules, periodic materials, and chemical reactions.
\end{abstract}

\maketitle

\section{introduction}

Free energy governs the direction, stability, and temperature dependence of physical, chemical, and biological processes. It determines reaction spontaneity, phase behavior, conformational equilibria, and molecular recognition, and therefore lies at the center of problems ranging from chemical reactions study \cite{peters2017reaction} to biomolecular function \cite{karplus2002molecular} and drug design \cite{reddy2001free}.
Despite its central importance, free energy remains notoriously difficult to compute because it depends on the partition function, an integral over a high-dimensional configuration space that is rarely tractable in realistic systems.

In practice, free energy is often inferred through free energy differences between thermodynamic states rather than computed absolutely. Classical approaches such as free energy perturbation (FEP) and thermodynamic integration (TI) have long provided the foundation for such calculations \cite{chipot2007free}. Both methods require samples from molecular dynamics (MD) or Markov chain Monte Carlo (MCMC). However, both MD and MCMC are limited by the sampling efficiency \cite{li2025unveiling,yu2026predicting}. MD requires appropriate time steps and is therefore expensive for long timescales, while MCMC often mixes slowly and struggles with rare transitions in high-dimensional or rugged landscapes.

Deep generative modeling has recently emerged as a transformative paradigm for directly sampling complex Boltzmann distributions \cite{li2018neural,wu2019solving, noe2019boltzmann}. By replacing the long sequential trajectories and inherently correlated sampling of conventional MD and MCMC methods with one-shot independent generation, these approaches bypass severe autocorrelation bottlenecks and accelerate the exploration of rare conformations\cite{schopmans2024conditional,asghar2024efficient,negedly2026correlation}. Beyond direct sampling, deep generative architectures are increasingly used for free energy calculations \cite{schebek2026assessing}. Key developments span discrete normalizing flows built upon invertible target mappings \cite{li2018neural, wirnsberger2020targeted, rizzi2021targeted, willow2023learned, schebek2024efficient, li2025deep, schebek2026assessing}, continuous normalizing flows integrated with flow matching objectives \cite{zhao2023bounding, hoffmann2026boltzmann}, and stochastic differential equation (SDE) driven diffusion models \cite{mate2024neural, he2026feat}.

While much progress has been made, scaling these methods to multiple thermal states remains computationally expensive. Traditional approaches rely on training discrete models or running independent simulations for each target temperature, incurring prohibitive computational costs when thermodynamic properties are required over broad temperature intervals. To overcome this limitation, recent work has begun to integrate temperature as an explicit conditioning variable into network architectures \cite{dibak2022temperature} or training frameworks \cite{li2025deep, li2026differentiable}. In particular, explicit temperature conditioning helps mitigate sampling bias and model collapse and significantly improves the accuracy of Boltzmann sampling. For instance, incorporating temperature as a conditioning factor enables the model to capture subtle phase transitions \cite{li2025deep}.

Here, we propose the learned temperature-differentiable free energy perturbation (TD-FEP) method. TD-FEP parameterizes the transformation between a reference and a target system as a temperature-continuous mapping and trains it in a  data-free, energy-based manner. This allows a single model to represent thermodynamic transformations over a range of temperatures, avoiding separate estimators at each condition. Except for one-shot generation of samples, TD-FEP yields a continuous representation of free energy with respect to temperature and enables differentiable estimation of related thermodynamic quantities. We validate the method on ethylene oxide, graphene, and the water–gas shift reaction, showing that it supports accurate free energy estimation while providing thermodynamic information across a continuous temperature range.

\section{Results and Discussion}

\subsection{Learned temperature-differentiable free energy perturbation (TD-FEP)}

\begin{figure}[htbp]
    \centering
    \includegraphics[width=0.8\linewidth]{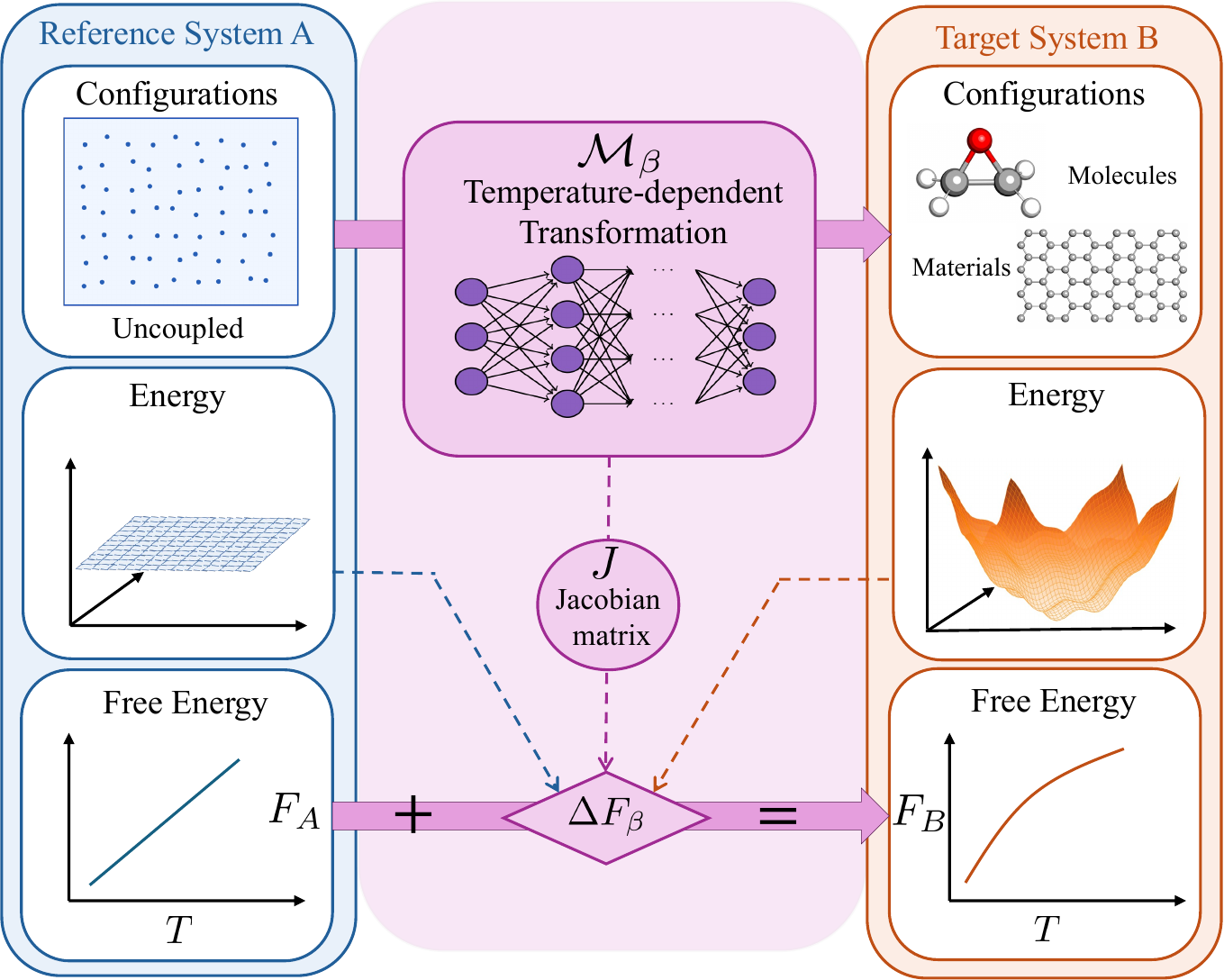}
    \caption{Schematic illustration of the learned temperature-differentiable free energy perturbation (TD-FEP) method. A temperature-dependent transformation $\mathcal{M}_\beta$ is employed to transform the configurations from an uncoupled reference system A
    into target system B of interest with the Boltzmann distribution. The free energy difference $\Delta F_\beta$ between system B and A is computed with their energy functions and the Jacobian matrix by $\mathcal{M}_\beta$. As the free energy $F_A$ of system A is tractable, the free energy $F_B$ of system B is calculated via $F_B = F_A +\Delta F_\beta$ }
    \label{fig:toc}
\end{figure}

In this section, we describe the details of the learned temperature-differentiable free energy estimation (TD-FEP) method. As illustrated in Fig.~\ref{fig:toc}, our framework involves the reference system A and the target system B, respectively. System A is designed as a tractable uncoupled system ensuring that the configurations are easy to sample and the free energy can be analytically computed (e.g., a uniform distribution). Let $U_A$ and $p_A$ denote the energy function and the probability distribution of system A, respectively. The target system B represents the physical system of interest, such as molecules (e.g., ethylene oxide) or crystalline materials (e.g., graphene). Let $U_B$ and $p_B$ denote the energy function and the probability distribution of system B, respectively. At given inverse temperature $\beta$, the samples from system A ($\bm x_A$) and system B ($\bm x_B$) follow the Boltzmann distribution:
\begin{equation}
    p_i(\bm x_i)
    = \frac{\exp[-\beta U_i(\bm x_i)]}{\mathcal{Z}_i},
    \qquad i=A,B.
\end{equation}
where $\mathcal{Z}_A$ and $\mathcal{Z}_B$ are the partition functions. The corresponding free energies are given by
\begin{align}
    F_i = -\frac{1}{\beta} \ln \mathcal{Z}_i, i=A,B
\end{align}

While $F_A$ is tractable by construction, computing $F_B$ directly is generally difficult as the integration over all configuration space is required to compute $\mathcal{Z}_B$.
Instead, we target the free energy difference $\Delta F_\beta$ between the two systems at an inverse temperature $\beta$, defined as
\begin{align}
    \Delta F_\beta = F_B - F_A = -\frac{1}{\beta} \ln \frac{\mathcal{Z}_B}{\mathcal{Z}_A}.
\end{align}
To achieve this, we employ a normalizing flow model as an invertible temperature-dependent transformation $\mathcal{M}_{\beta}^{\theta}$ to bridge two systems, following our previous variational temperature-differentiable (VaTD) method \cite{li2025deep}.
Using change-of-variable formula, the parameterized probability distribution, $q_{\theta}$, denotes the variational density of system A.

Using Kullback-Leibler divergence (KLD) \cite{kullback1951information}, we can quantify the difference between parameterized $q_{\theta}$ and $p_B$ of target system B, as
\begin{equation}
    \begin{aligned}
        D_{KL}(q_\theta||p_B) = & \underset{\bm x \sim q_\theta}{\mathbb{E}}\left[\log q_\theta(\bm x) + \beta U_B(\bm x)\right] +\log \mathcal{Z}_B \\
        = &  \beta \left(\mathbb{E}_{\bm x\sim p_A} \left[ U_B(\mathcal{M}_\beta^\theta(\bm x)) - U_A(\bm x) -\beta^{-1}\log|\det J| \right] -\Delta F_\beta  \right) \geq 0
    \end{aligned}
    \label{eq:DKL}
\end{equation}

We note the former estimation as the generalized work from reference system A to target system B (forward direction) \cite{hahn2009using,wirnsberger2020targeted},
$W_F(\bm x_A; \beta) =U_B\!\left(\mathcal{M}_\beta^{\theta}(\mathbf{x})\right) - U_A(\bm x) -\beta^{-1} \ln \left|\det \mathbf{J}\right|$, which includes the mean energy difference $U_B -U_A$ and entropy difference $\ln|\text{det} J|$.

At given temperature, $\frac{1}{\beta}  D_{KL}(q_\theta(\cdot, \beta)||p_B)$ measures this estimated free energy gap between $\mathbb{E}_{\bm x_A\sim p_A}[W_F(\bm x_A;\beta)]$ and $\Delta F_\beta$. We minimize this gap over a continuous temperature range as the loss function:
\begin{align}
    \mathcal{L}
    &= \underset{\substack{\beta \sim \mathcal{U} \\ \mathbf{x_A} \sim p_A}}{\mathbb{E}} \left[\beta W_F(\bm x_A; \beta)\right]
\end{align}

As shown by \cite{jarzynski2002targeted},
$\Delta F_\beta$ and $W_F$ satisfy the identity:
\begin{align}
    \Delta F_\beta = -\frac{1}{\beta}\log \mathbb{E}_{\bm x_A\sim p_A} \left[\exp(-\beta W_F)\right]
    \label{eq:tFEPf}
\end{align}
When the mapping is sufficiently accurate and $q_\theta$ aligns well with $p_B$,
$\mathbb{E}_{x \sim p_A}[W_F ]$ yields consistent
free energy estimates with Eq.~\ref{eq:tFEPf}. In practice, however, if the mapping is not accurate enough,
$\mathbb{E}_{x \sim p_A}[ W_F ]$ provides only an upper bound that can become loose.

Since our transformation is continuous with respect to temperature, thermodynamic quantities can be obtained by differentiating the estimated free energy. For example, the average energy under the target Boltzmann distribution can be expressed as:
 \begin{align}
     \langle U \rangle_B = \frac{\partial (\beta F_B)}{\partial \beta} = \frac{\partial}{\partial \beta} \beta \left[ F_A +\Delta F_{\beta} \right]
 \end{align}

\subsection{2D Gaussian-mixture system}

\begin{figure}[htbp]
    \centering
    \includegraphics[width=\textwidth]{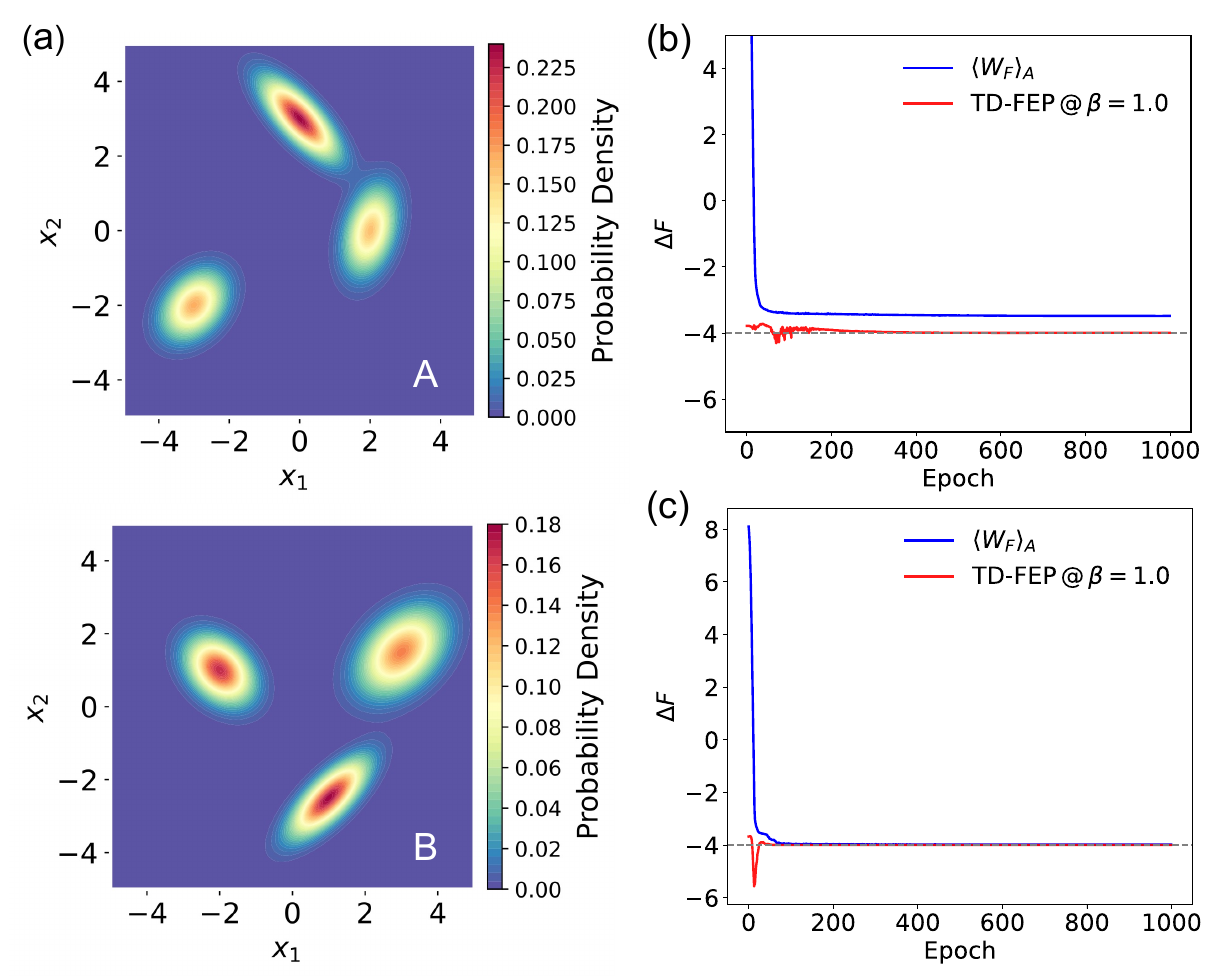}
    \caption{Calculation of $\Delta F_{BA}$ on a 2D energy landscape. (a) Probability density contours for system A (top) and system B (bottom). (b) Convergence of estimated free energy difference versus training epochs. The ground truth $\Delta F_{BA}$ is $-4.0$. Estimations are shown for $\langle W_F \rangle_A$ (blue) and
    our proposed TD-FEP method at $\beta=1.0$ (red).
    (c) Convergence profile with a better mapping demonstrating improved estimation of $\langle W_F \rangle_A$.}
    \label{fig:2dgs}
\end{figure}

To validate the proposed temperature-differentiable framework, we first evaluated the free energy difference estimation between two triple-well potential systems, $A$ and $B$, as illustrated by the probability density contours in Fig.~\ref{fig:2dgs} a. Both systems were constructed using truncated Gaussian mixtures to represent complex, multi-modal energy landscapes.
By adjusting the reference energy zero point, the free energy difference between systems is set as -4.0 exactly. The inverse temperature $\beta$ is equivalent to 1.0 in  this case. The invertible mapping $\mathcal{M}_{\beta}^{\theta}$ was parameterized by a piecewise linear normalizing flow \cite{durkan2019neural}.

Fig.~\ref{fig:2dgs} b and c show the convergence curves of free energy difference computed with $\langle W_F\rangle$ and forward free energy estimator by Eq.~(\ref{eq:tFEPf}).
As shown in Fig.~\ref{fig:2dgs} b, when the model is not sufficiently trained and the mapping is not perfect, the $\langle W_F\rangle_A$ gives the upper bound of the free energy difference, $\Delta F_{BA}$, from a distance above, while TD-FEP maintains a closer alignment.
With a model of sufficient capacity and well-trained mapping, as shown in Fig.~\ref{fig:2dgs} c, both estimates converge to the theoretical value $-4.0$.
This is consistent with our expectation. When the mapping is accurate, the parameterized distribution $q_\theta$ aligns well with $p_B$, and the standard deviation of $W_F$ becomes small. In this regime, the exponential expectation in Eq.~\ref{eq:tFEPf} can be approximated as $\mathbb{E}_{p_A}[e^{-\beta W_F}] \approx e^{-\beta \langle W_F\rangle_A}$, so that Eq.~\ref{eq:tFEPf} reduces to $\Delta F_\beta \approx \langle W_F\rangle_A$. More generally, by Jensen's inequality, $\langle W_F\rangle_A$ always provides an upper bound to $\Delta F_\beta$ in Eq.~\ref{eq:tFEPf}, with equality only in the limit of vanishing fluctuations in $W_F$.
The training details and parameter settings can be found in our SI.

\subsection{Ethylene oxide molecular system}

\begin{figure}[htbp]
    \centering
    \includegraphics[width=\textwidth]{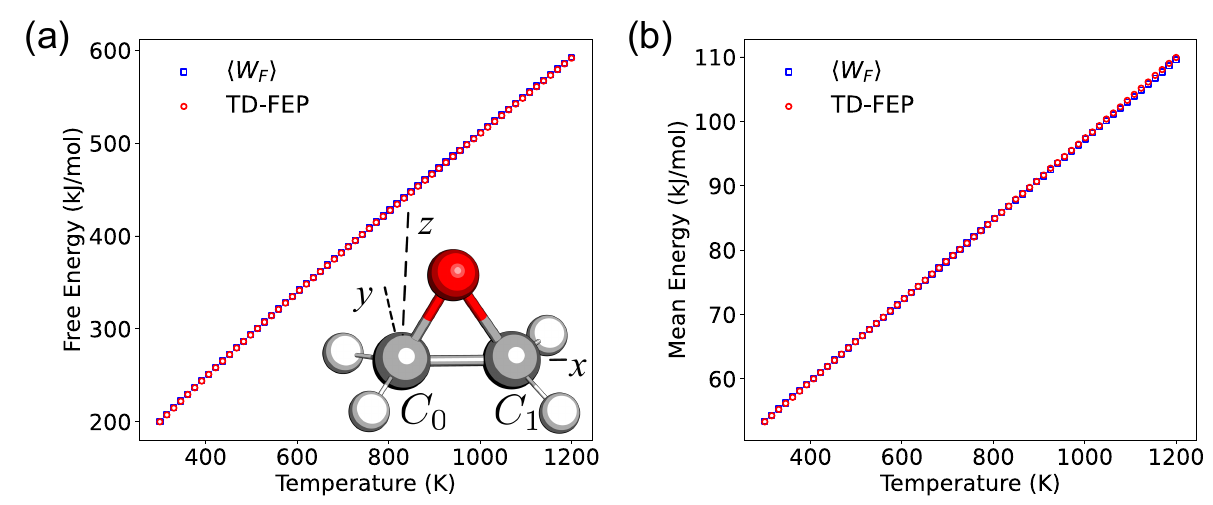}
    \caption{Temperature dependence of the configurational free energy and mean energy for ethylene oxide. (a) Configurational free energy $F$ calculated via
    the expectation of  work $\langle W_F \rangle$ (blue),
    and learned temperature-differentiable free energy perturbation (TD-FEP) method.
    An inset illustrates the ethylene oxide molecule within the Cartesian coordinate system. (b) Mean energy as a function of temperature, comparing the derivative computation of $\langle W_F \rangle$ (blue) and the TD-FEP results (red).
    }
    \label{fig:C2H4O}
\end{figure}

We further apply the proposed TD-FEP method to a realistic molecular system: ethylene oxide ($\text{C}_2\text{H}_4\text{O}$), which consists of 7 atoms and 15 vibrational degrees of freedom. This molecule has been extensively studied in previous computational works \cite{komornicki1983ab, lowe1986scaled} and is frequently utilized as a benchmark for methodological illustrations \cite{begue2006single, thomas2015using, zhang2024neural}. In this study, we employ the GROMOS 54A7 force field \cite{schmid2011definition} to model the molecule, with parameters generated via the Automated Topology Builder (ATB) \cite{malde2011automated, stroet2018automated}.

As shown in the inset of Fig.~\ref{fig:C2H4O} a, a local coordinate system is defined to eliminate global translational and rotational degrees of freedom: carbon atom $C_0$ is fixed at the origin, the second carbon atom $C_1$ is constrained to the $x$-axis, and the oxygen atom is restricted to the $xz$-plane. This construction reduces the system to a 15-dimensional configurational space.

The configurational free energy $F$ of the system is computed using the expectation of work $\langle W_F \rangle$, and the proposed TD-FEP method.
As shown in Fig.~\ref{fig:C2H4O} a, the two methods agree well with each other, indicating a well-trained transformation.
The mean energy $E$ is derived from the derivative of the free energy with respect to $\beta$, with results presented in Fig.~\ref{fig:C2H4O} b.
These results align with each other.
Furthermore, these profiles of the mean energy coincide with the theoretical equipartition theorem of statistical physics.
\subsection{Graphene crystal system}
\begin{figure}[htbp]
    \centering
    \includegraphics[width=\textwidth]{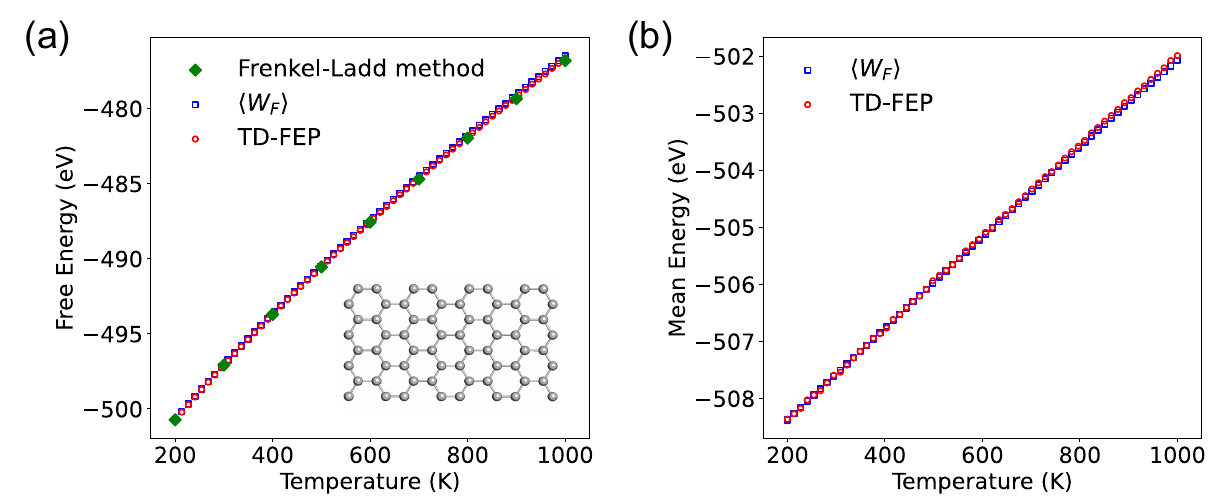}
    \caption{Temperature dependence of the configurational free energy and mean energy for graphene with 64 atoms. (a) Configurational free energy $F$ calculated via Frenkel-Ladd method (green), the expectation of  work $\langle W_F \rangle$ (blue), and the learned temperature-differentiable free energy perturbation (TD-FEP) method (red). An inset illustrates the configuration of graphene. (b) Mean energy as a function of temperature, comparing the derivative computation of $\langle W_F \rangle$ (blue) and the TD-FEP results (red). }
    \label{fig:graphene}
\end{figure}

The proposed method can be further extended to crystals with periodic boundary conditions (PBCs). As a case study, we employ a 64-atom graphene system in a rectangular cell under PBCs. To eliminate translational invariance, one carbon atom is fixed while the remaining atoms are confined within a cubic region ($1.0\text{\AA}^3$) relative to the fixed atom. These cubic regions are centered at the equilibrium position of the corresponding graphene atoms.
The well-known Tersoff   \cite{tersoff1989modeling, ouyang2018nanoserpents, ouyang2019mechanical} potential is employed as the force field to model the graphene system. For reference, we use LAMMPS \cite{thompson2022lammps} with the Frenkel–Ladd method \cite{freitas2016nonequilibrium} to compute the configurational free energy of the graphene system. In the Frenkel–Ladd method, the Einstein crystal is chosen as the reference system owing to its analytically known partition function.
For a fair comparison, we applied the same translational constraint as in TD-FEP
(see the SI for details).

The estimated free energy as a function of temperature is shown in Fig.~\ref{fig:graphene}a. Our proposed TD-FEP method performs well in a continuous temperature range. Both $\langle W_F \rangle$ and TD-FEP give good estimates in the low-temperature region. As the temperature increases, the estimates from $\langle W_F \rangle$ deviate slightly, whereas our TD-FEP method still agrees well with the Frenkel--Ladd method. The mean energy is then calculated via the derivative of the estimated free energy as shown in Fig.~\ref{fig:graphene}b. The profile of the mean energy again coincides with the equipartition theorem of statistical physics.
We thus estimate, in future developments, the proposed TD-FEP method could be integrated with cutting-edge neural-network models to implement the continuous-temperature-range free-energy calculation for large-scale materials under PBC, such as~\cite{schebek2026scalable}.

\subsection{Water-gas shift reaction}

\begin{figure}[htbp]
    \centering
    \includegraphics[width=\textwidth]{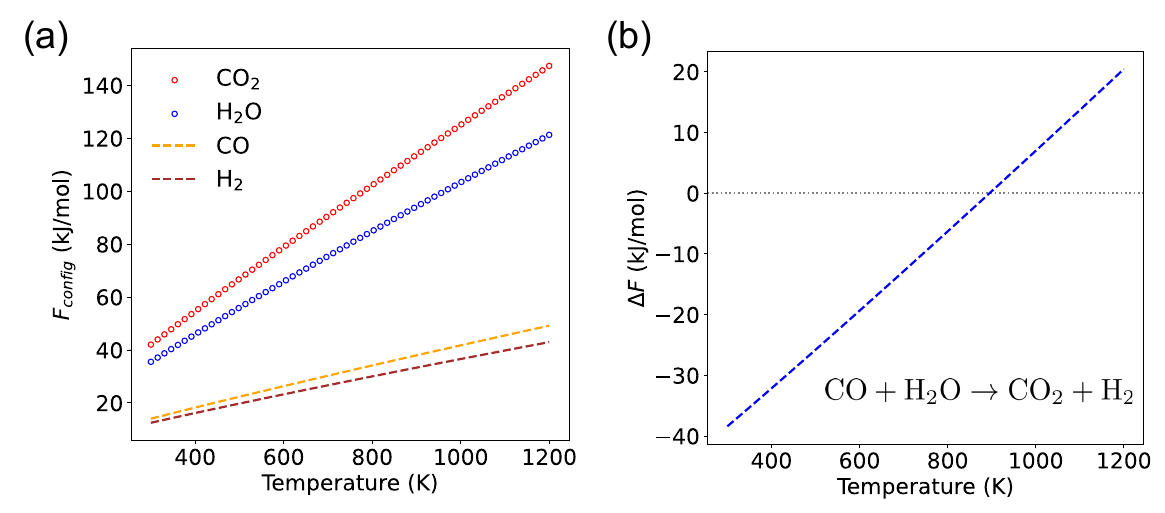}
    \caption{Free energy calculation of the water-gas shift reaction ($\text{CO + H}_2\text{O} \rightleftharpoons \text{CO}_2 + \text{H}_2$) in the dilute limit. (a) Configurational free energy ($F_{\text{config}}$) by TD-FEP method as a function of temperature for different species: CO$_2$ (red open circles), H$_2$O (blue open circles), CO (yellow dashed line), and H$_2$ (brown dashed line). (b) Temperature dependence of the total Helmholtz free energy difference ($\Delta F$).}
    \label{fig:wgs}
\end{figure}

To further demonstrate the applicability of our method to chemical reactions, we consider the well-known water-gas shift reaction in the dilute limit,
\begin{align}
\text{CO} + \text{H}_2\text{O} \rightleftharpoons \text{CO}_2 + \text{H}_2,
\end{align}
 where the Gibbs free energy change is the key thermodynamic quantity characterizing this process.
Formally, the Gibbs free energy change, $\Delta G$, can be defined as
\begin{align}
\Delta G = \Delta F + P \Delta V
\end{align}
where $\Delta F$ is the Helmholtz free energy change, $P$ is the pressure and $\Delta V$ is the volume change. In ideal-gas situation, as the water-gas shift reaction preserves the total number of molecules, the $\Delta V$ approaches zero thus vanishing the $P \Delta V$ contribution which allows us to focus on the Helmholtz free energy change, $\Delta F$. This term can be decomposed into several contributions:
\begin{align}
\Delta F = \Delta E_e + \Delta F_{\text{trans}} + \Delta F_{\text{rot}} + \Delta F_{\text{config}}
\end{align}
where $\Delta E_e$ represents the electronic energy difference, $\Delta F_{\text{trans}}$, $\Delta F_{\text{rot}}$, and $\Delta F_{\text{config}}$ denote the translational, rotational, and configurational contributions, respectively.

The $\Delta F_{\text{config}}$, can be computed using our  TD-FEP  method.
In principle, the partition function should formally include the kinetic energy contribution, along with the constant factor, $h^{3N}$, where $h$ is Planck's constant and $N$ is the number of atoms.
This accounts for the quantization of phase space and ensures the free energy remains independent of the choice of coordinate units. However, in the context of free energy differences between reactants and products, these two terms can be safely omitted as they have the equal contribution to the reactants and products.
The details of the computing $\Delta F_{\text{trans}}$, $\Delta F_{\text{rot}}$ are provided in the SI.
$\Delta E_e$ is then computed from the density functional theory (DFT) using PySCF \cite{sun2015libcint, sun2018pyscf, sun2020recent},
which obtains $\Delta E_e$ as $-55.72$ KJ/mol.
The details about the DFT computation are also provided in the SI.

Fig.~\ref{fig:wgs} a shows the computed configurational free energy ($F_{\text{config}}$) for each chemical species
involved in the water-gas shift reaction. The $F_{\text{config}}$ for CO$_2$ and H$_2$O were determined using the proposed TD-FEP method with the mass center fixed to eliminate translational invariance.
As illustrated,
$F_{\text{config}}$ contribution from CO$_2$ is above that from H$_2$O
with widening gaps at elevated temperatures.
This trend is physically consistent with the fact that CO$_2$, as a triatomic linear molecule, possesses more vibrational degrees of freedom (four) compared to the nonlinear H$_2$O (three). Since each vibrational mode contributes an entropic term, the additional mode in CO$_2$ gives it a larger vibrational entropy. As temperature increases, this entropic term becomes more pronounced, leading to a widening free energy gap.
For the diatomic molecules CO and H$_2$, $F_{\text{config}}$ was calculated via numerical integration, providing an exact treatment for these simpler systems. As above, we implement the GROMOS 54A7 force field \cite{schmid2011definition} to model the potential energy surface of molecules.

Fig.~\ref{fig:wgs} b shows the total Helmholtz free energy change, $\Delta F$, from all above contributions with respect to temperature. At the low temperature region, $\Delta F$ is negative and the reaction is exothermic.
As the temperature increases, molecules have more kinetic energy and are more likely to overcome the free-energy barrier, resulting in accelerating reaction rates.
However, $\Delta F$ gradually approaches zero with rising temperature. When the temperature exceeds about $900 K$, $\Delta F$ turns from negative into positive, indicating that the reaction becomes thermodynamically unfavorable in the forward direction and tends to proceed in the reverse direction.

\section{Conclusion}
In conclusion, we have developed the learned temperature-differentiable free energy perturbation (TD-FEP) method that enables the continuous estimation of free energy across a wide temperature range within a single model. By parameterizing the transformation between thermodynamic states as a temperature-continuous mapping, we avoid
the redundant sampling at multiple thermal states. Our results on the ethylene oxide, graphene and water-gas shift reaction demonstrate the accuracy of proposed method.
Our temperature-continuous generative framework offers a path toward fully automated and differentiable thermodynamic modeling of complex chemical processes.

\section{Method}
A normalizing flow \cite{papamakarios2021normalizing, rezende2015variational} parameterizes a target probability distribution by composing a series of invertible transformations (i.e., bijections) atop a simple prior distribution $q(\cdot)$. By applying the change-of-variables rule, the log-probability of an observation $\boldsymbol{x}$ under the resulting flow-based model $q_{\theta}$ can be expressed as
\[
\ln q_{\theta}(\boldsymbol{x})
= \ln q\left(\mathcal{M}_{\theta}^{-1}(\boldsymbol{x})\right)
+ \ln \left| \det \frac{\partial \mathcal{M}_{\theta}^{-1}(\boldsymbol{x})}{\partial \boldsymbol{x}} \right|,
\]
where $\mathcal{M}_{\theta}(\cdot)$ denotes the parameterized bijective mapping and $\mathcal{M}_{\theta}^{-1}(\cdot)$ represents its inverse transformation. In practice, $\mathcal{M}_{\theta}$ is realized via a sequence of coupling layers. Specifically, each coupling layer partitions the input variables into two disjoint groups: one group is kept unchanged and is used to condition a neural network, whose outputs parameterize element-wise, invertible transformations applied to the other group.

In spline flows \cite{durkan2019neural,durkan2019cubic}, each coupling layer employs an element-wise monotonic piecewise spline. For $K$ bins on $[0,1]$, the neural network outputs $K$ bin widths, $K$ bin heights, and two boundary derivatives for each transformed variable. These parameters collectively define a continuously differentiable monotone piecewise spline function.

\section{Data Availability}

The data that support this study are available upon request from the authors.

\section{Code Availability}
The code is available at https://github.com/phyzyw/TD-FEP1

\section{Acknowledgements}
S.-H.L. and D.P. acknowledge support from Hong Kong Research Grants Council (RGC) (GRF-16302423).
D.P. acknowledges support from Hong Kong RGC (GRF-16301723, GRF-16310225),
National Natural Science Foundation of China/RGC Joint Research Scheme (N\_HKUST664/24).
Part of this work was carried out using computational resources from the National Supercomputer Center in Guangzhou, China, the X-GPU cluster supported by the RGC Collaborative Research Fund C6021-19EF, and the High Performance Computing facilities, HKUST SuperPOD, provided by ITSO, The Hong Kong University of Science and Technology (HKUST).

\section{Competing Interests}
The authors declare no competing interests.

\bibliography{ref}

\begin{thebibliography}{10}
\expandafter\ifx\csname url\endcsname\relax
  \def\url#1{\texttt{#1}}\fi
\expandafter\ifx\csname urlprefix\endcsname\relax\def\urlprefix{URL }\fi
\providecommand{\bibinfo}[2]{#2}
\providecommand{\eprint}[2][]{\url{#2}}

\bibitem{peters2017reaction}
\bibinfo{author}{Peters, B.}
\newblock \emph{\bibinfo{title}{Reaction rate theory and rare events}}
  (\bibinfo{publisher}{Elsevier}, \bibinfo{year}{2017}).

\bibitem{karplus2002molecular}
\bibinfo{author}{Karplus, M.} \& \bibinfo{author}{McCammon, J.~A.}
\newblock \bibinfo{title}{Molecular dynamics simulations of biomolecules}.
\newblock \emph{\bibinfo{journal}{Nature structural biology}}
  \textbf{\bibinfo{volume}{9}}, \bibinfo{pages}{646--652}
  (\bibinfo{year}{2002}).

\bibitem{reddy2001free}
\bibinfo{author}{Reddy, M.~R.} \& \bibinfo{author}{Erion, M.~D.}
\newblock \emph{\bibinfo{title}{Free energy calculations in rational drug
  design}} (\bibinfo{publisher}{Springer Science \& Business Media},
  \bibinfo{year}{2001}).

\bibitem{chipot2007free}
\bibinfo{author}{Chipot, C.} \& \bibinfo{author}{Pohorille, A.}
\newblock \emph{\bibinfo{title}{Free energy calculations}},
  vol.~\bibinfo{volume}{86} (\bibinfo{publisher}{Springer},
  \bibinfo{year}{2007}).

\bibitem{li2025unveiling}
\bibinfo{author}{Li, C.}, \bibinfo{author}{Yao, Y.} \& \bibinfo{author}{Pan,
  D.}
\newblock \bibinfo{title}{Unveiling hidden reaction kinetics of carbon dioxide
  in supercritical aqueous solutions}.
\newblock \emph{\bibinfo{journal}{Proceedings of the National Academy of
  Sciences}} \textbf{\bibinfo{volume}{122}}, \bibinfo{pages}{e2406356121}
  (\bibinfo{year}{2025}).

\bibitem{yu2026predicting}
\bibinfo{author}{Yu, J.}, \bibinfo{author}{Li, S.-H.} \& \bibinfo{author}{Pan,
  D.}
\newblock \bibinfo{title}{Predicting solvation free energies of molecules and
  ions via first-principles and machine-learning molecular dynamics}.
\newblock \emph{\bibinfo{journal}{Journal of Chemical Theory and Computation}}
  \textbf{\bibinfo{volume}{22}}, \bibinfo{pages}{6377} (\bibinfo{year}{2026}).

\bibitem{li2018neural}
\bibinfo{author}{Li, S.-H.} \& \bibinfo{author}{Wang, L.}
\newblock \bibinfo{title}{Neural network renormalization group}.
\newblock \emph{\bibinfo{journal}{Physical review letters}}
  \textbf{\bibinfo{volume}{121}}, \bibinfo{pages}{260601}
  (\bibinfo{year}{2018}).

\bibitem{wu2019solving}
\bibinfo{author}{Wu, D.}, \bibinfo{author}{Wang, L.} \& \bibinfo{author}{Zhang,
  P.}
\newblock \bibinfo{title}{Solving statistical mechanics using variational
  autoregressive networks}.
\newblock \emph{\bibinfo{journal}{Physical review letters}}
  \textbf{\bibinfo{volume}{122}}, \bibinfo{pages}{080602}
  (\bibinfo{year}{2019}).

\bibitem{noe2019boltzmann}
\bibinfo{author}{No{\'e}, F.}, \bibinfo{author}{Olsson, S.},
  \bibinfo{author}{K{\"o}hler, J.} \& \bibinfo{author}{Wu, H.}
\newblock \bibinfo{title}{Boltzmann generators: Sampling equilibrium states of
  many-body systems with deep learning}.
\newblock \emph{\bibinfo{journal}{Science}} \textbf{\bibinfo{volume}{365}},
  \bibinfo{pages}{eaaw1147} (\bibinfo{year}{2019}).

\bibitem{schopmans2024conditional}
\bibinfo{author}{Schopmans, H.} \& \bibinfo{author}{Friederich, P.}
\newblock \bibinfo{title}{Conditional normalizing flows for active learning of
  coarse-grained molecular representations}.
\newblock \emph{\bibinfo{journal}{arXiv preprint arXiv:2402.01195}}
  (\bibinfo{year}{2024}).

\bibitem{asghar2024efficient}
\bibinfo{author}{Asghar, S.}, \bibinfo{author}{Pei, Q.-X.},
  \bibinfo{author}{Volpe, G.} \& \bibinfo{author}{Ni, R.}
\newblock \bibinfo{title}{Efficient rare event sampling with unsupervised
  normalizing flows}.
\newblock \emph{\bibinfo{journal}{Nature Machine Intelligence}}
  \textbf{\bibinfo{volume}{6}}, \bibinfo{pages}{1370--1381}
  (\bibinfo{year}{2024}).

\bibitem{negedly2026correlation}
\bibinfo{author}{Negedly, M.}, \bibinfo{author}{Falkner, S.},
  \bibinfo{author}{Coretti, A.} \& \bibinfo{author}{Dellago, C.}
\newblock \bibinfo{title}{Correlation-free transition path sampling through
  shooting point generation guided by committor learning}.
\newblock \emph{\bibinfo{journal}{arXiv preprint arXiv:2609.20461}}
  (\bibinfo{year}{2026}).

\bibitem{schebek2026assessing}
\bibinfo{author}{Schebek, M.} \emph{et~al.}
\newblock \bibinfo{title}{Assessing generative modeling approaches for free
  energy estimates in condensed matter}.
\newblock \emph{\bibinfo{journal}{The Journal of Chemical Physics}}
  \textbf{\bibinfo{volume}{164}} (\bibinfo{year}{2026}).

\bibitem{wirnsberger2020targeted}
\bibinfo{author}{Wirnsberger, P.} \emph{et~al.}
\newblock \bibinfo{title}{Targeted free energy estimation via learned
  mappings}.
\newblock \emph{\bibinfo{journal}{The Journal of Chemical Physics}}
  \textbf{\bibinfo{volume}{153}} (\bibinfo{year}{2020}).

\bibitem{rizzi2021targeted}
\bibinfo{author}{Rizzi, A.}, \bibinfo{author}{Carloni, P.} \&
  \bibinfo{author}{Parrinello, M.}
\newblock \bibinfo{title}{Targeted free energy perturbation revisited: Accurate
  free energies from mapped reference potentials}.
\newblock \emph{\bibinfo{journal}{The journal of physical chemistry letters}}
  \textbf{\bibinfo{volume}{12}}, \bibinfo{pages}{9449--9454}
  (\bibinfo{year}{2021}).

\bibitem{willow2023learned}
\bibinfo{author}{Willow, S.~Y.}, \bibinfo{author}{Kang, L.} \&
  \bibinfo{author}{Minh, D.~D.}
\newblock \bibinfo{title}{Learned mappings for targeted free energy
  perturbation between peptide conformations}.
\newblock \emph{\bibinfo{journal}{The Journal of Chemical Physics}}
  \textbf{\bibinfo{volume}{159}} (\bibinfo{year}{2023}).

\bibitem{schebek2024efficient}
\bibinfo{author}{Schebek, M.}, \bibinfo{author}{Invernizzi, M.},
  \bibinfo{author}{No{\'e}, F.} \& \bibinfo{author}{Rogal, J.}
\newblock \bibinfo{title}{Efficient mapping of phase diagrams with conditional
  boltzmann generators}.
\newblock \emph{\bibinfo{journal}{Machine Learning: Science and Technology}}
  \textbf{\bibinfo{volume}{5}}, \bibinfo{pages}{045045} (\bibinfo{year}{2024}).

\bibitem{li2025deep}
\bibinfo{author}{Li, S.-H.}, \bibinfo{author}{Zhang, Y.-W.} \&
  \bibinfo{author}{Pan, D.}
\newblock \bibinfo{title}{Deep generative modeling of the canonical ensemble
  with differentiable thermal properties}.
\newblock \emph{\bibinfo{journal}{Physical Review Letters}}
  \textbf{\bibinfo{volume}{135}}, \bibinfo{pages}{027301}
  (\bibinfo{year}{2025}).

\bibitem{zhao2023bounding}
\bibinfo{author}{Zhao, L.} \& \bibinfo{author}{Wang, L.}
\newblock \bibinfo{title}{Bounding free energy difference with flow matching}.
\newblock \emph{\bibinfo{journal}{Chinese Physics Letters}}
  \textbf{\bibinfo{volume}{40}}, \bibinfo{pages}{120201}
  (\bibinfo{year}{2023}).

\bibitem{hoffmann2026boltzmann}
\bibinfo{author}{Hoffmann, E.}, \bibinfo{author}{Schebek, M.},
  \bibinfo{author}{Klein, L.}, \bibinfo{author}{No{\'e}, F.} \&
  \bibinfo{author}{Rogal, J.}
\newblock \bibinfo{title}{Boltzmann generators for condensed matter via
  riemannian flow matching}.
\newblock \emph{\bibinfo{journal}{arXiv preprint arXiv:2602.18482}}
  (\bibinfo{year}{2026}).

\bibitem{mate2024neural}
\bibinfo{author}{Mate, B.}, \bibinfo{author}{Fleuret, F.} \&
  \bibinfo{author}{Bereau, T.}
\newblock \bibinfo{title}{Neural thermodynamic integration: Free energies from
  energy-based diffusion models}.
\newblock \emph{\bibinfo{journal}{The Journal of Physical Chemistry Letters}}
  \textbf{\bibinfo{volume}{15}}, \bibinfo{pages}{11395--11404}
  (\bibinfo{year}{2024}).

\bibitem{he2026feat}
\bibinfo{author}{Du, Y.} \emph{et~al.}
\newblock \bibinfo{title}{Feat: Free energy estimators with adaptive
  transport}.
\newblock \emph{\bibinfo{journal}{Advances in Neural Information Processing
  Systems}} \textbf{\bibinfo{volume}{38}}, \bibinfo{pages}{142130--142167}
  (\bibinfo{year}{2026}).

\bibitem{dibak2022temperature}
\bibinfo{author}{Dibak, M.}, \bibinfo{author}{Klein, L.},
  \bibinfo{author}{Kr{\"a}mer, A.} \& \bibinfo{author}{No{\'e}, F.}
\newblock \bibinfo{title}{Temperature steerable flows and boltzmann
  generators}.
\newblock \emph{\bibinfo{journal}{Physical Review Research}}
  \textbf{\bibinfo{volume}{4}}, \bibinfo{pages}{L042005}
  (\bibinfo{year}{2022}).

\bibitem{li2026differentiable}
\bibinfo{author}{Li, S.-H.}, \bibinfo{author}{Chen, C.},
  \bibinfo{author}{Zhang, Y.-W.} \& \bibinfo{author}{Pan, D.}
\newblock \bibinfo{title}{Differentiable free energy surface: a variational
  approach to directly observing rare events using generative deep-learning
  models}.
\newblock \emph{\bibinfo{journal}{arXiv preprint arXiv:2604.09769}}
  (\bibinfo{year}{2026}).

\bibitem{kullback1951information}
\bibinfo{author}{Kullback, S.} \& \bibinfo{author}{Leibler, R.~A.}
\newblock \bibinfo{title}{On information and sufficiency}.
\newblock \emph{\bibinfo{journal}{The annals of mathematical statistics}}
  \textbf{\bibinfo{volume}{22}}, \bibinfo{pages}{79--86}
  (\bibinfo{year}{1951}).

\bibitem{hahn2009using}
\bibinfo{author}{Hahn, A.~M.} \& \bibinfo{author}{Then, H.}
\newblock \bibinfo{title}{Using bijective maps to improve free-energy
  estimates}.
\newblock \emph{\bibinfo{journal}{Physical Review E—Statistical, Nonlinear,
  and Soft Matter Physics}} \textbf{\bibinfo{volume}{79}},
  \bibinfo{pages}{011113} (\bibinfo{year}{2009}).

\bibitem{jarzynski2002targeted}
\bibinfo{author}{Jarzynski, C.}
\newblock \bibinfo{title}{Targeted free energy perturbation}.
\newblock \emph{\bibinfo{journal}{Physical Review E}}
  \textbf{\bibinfo{volume}{65}}, \bibinfo{pages}{046122}
  (\bibinfo{year}{2002}).

\bibitem{durkan2019neural}
\bibinfo{author}{Durkan, C.}, \bibinfo{author}{Bekasov, A.},
  \bibinfo{author}{Murray, I.} \& \bibinfo{author}{Papamakarios, G.}
\newblock \bibinfo{title}{Neural spline flows}.
\newblock \emph{\bibinfo{journal}{Advances in neural information processing
  systems}} \textbf{\bibinfo{volume}{32}} (\bibinfo{year}{2019}).

\bibitem{komornicki1983ab}
\bibinfo{author}{Komornicki, A.}, \bibinfo{author}{Pauzat, F.} \&
  \bibinfo{author}{Ellinger, Y.}
\newblock \bibinfo{title}{Ab initio prediction of structures, force constants
  and vibrational frequencies. saturated three-membered rings cyclopropane,
  ethylene oxide, and ethyleneimine}.
\newblock \emph{\bibinfo{journal}{The Journal of Physical Chemistry}}
  \textbf{\bibinfo{volume}{87}}, \bibinfo{pages}{3847--3857}
  (\bibinfo{year}{1983}).

\bibitem{lowe1986scaled}
\bibinfo{author}{Lowe, M.~A.}, \bibinfo{author}{Alper, J.~S.},
  \bibinfo{author}{Kawiecki, R.} \& \bibinfo{author}{Stephens, P.}
\newblock \bibinfo{title}{Scaled ab initio force fields for ethylene oxide and
  propylene oxide}.
\newblock \emph{\bibinfo{journal}{The Journal of Physical Chemistry}}
  \textbf{\bibinfo{volume}{90}}, \bibinfo{pages}{41--50}
  (\bibinfo{year}{1986}).

\bibitem{begue2006single}
\bibinfo{author}{Begue, D.}, \bibinfo{author}{Gohaud, N.},
  \bibinfo{author}{Brown, R.} \& \bibinfo{author}{Pouchan, C.}
\newblock \bibinfo{title}{An single program multiple data strategy for
  calculation of anharmonic vibrations}.
\newblock \emph{\bibinfo{journal}{Journal of mathematical chemistry}}
  \textbf{\bibinfo{volume}{40}}, \bibinfo{pages}{197--211}
  (\bibinfo{year}{2006}).

\bibitem{thomas2015using}
\bibinfo{author}{Thomas, P.~S.} \& \bibinfo{author}{Carrington~Jr, T.}
\newblock \bibinfo{title}{Using nested contractions and a hierarchical tensor
  format to compute vibrational spectra of molecules with seven atoms}.
\newblock \emph{\bibinfo{journal}{The Journal of Physical Chemistry A}}
  \textbf{\bibinfo{volume}{119}}, \bibinfo{pages}{13074--13091}
  (\bibinfo{year}{2015}).

\bibitem{zhang2024neural}
\bibinfo{author}{Zhang, Q.}, \bibinfo{author}{Wang, R.-S.} \&
  \bibinfo{author}{Wang, L.}
\newblock \bibinfo{title}{Neural canonical transformations for vibrational
  spectra of molecules}.
\newblock \emph{\bibinfo{journal}{The Journal of Chemical Physics}}
  \textbf{\bibinfo{volume}{161}} (\bibinfo{year}{2024}).

\bibitem{schmid2011definition}
\bibinfo{author}{Schmid, N.} \emph{et~al.}
\newblock \bibinfo{title}{Definition and testing of the gromos force-field
  versions 54a7 and 54b7}.
\newblock \emph{\bibinfo{journal}{European biophysics journal}}
  \textbf{\bibinfo{volume}{40}}, \bibinfo{pages}{843--856}
  (\bibinfo{year}{2011}).

\bibitem{malde2011automated}
\bibinfo{author}{Malde, A.~K.} \emph{et~al.}
\newblock \bibinfo{title}{An automated force field topology builder (atb) and
  repository: version 1.0}.
\newblock \emph{\bibinfo{journal}{Journal of chemical theory and computation}}
  \textbf{\bibinfo{volume}{7}}, \bibinfo{pages}{4026--4037}
  (\bibinfo{year}{2011}).

\bibitem{stroet2018automated}
\bibinfo{author}{Stroet, M.} \emph{et~al.}
\newblock \bibinfo{title}{Automated topology builder version 3.0: Prediction of
  solvation free enthalpies in water and hexane}.
\newblock \emph{\bibinfo{journal}{Journal of chemical theory and computation}}
  \textbf{\bibinfo{volume}{14}}, \bibinfo{pages}{5834--5845}
  (\bibinfo{year}{2018}).

\bibitem{tersoff1989modeling}
\bibinfo{author}{Tersoff, J.}
\newblock \bibinfo{title}{Modeling solid-state chemistry: Interatomic
  potentials for multicomponent systems}.
\newblock \emph{\bibinfo{journal}{Physical review B}}
  \textbf{\bibinfo{volume}{39}}, \bibinfo{pages}{5566} (\bibinfo{year}{1989}).

\bibitem{ouyang2018nanoserpents}
\bibinfo{author}{Ouyang, W.}, \bibinfo{author}{Mandelli, D.},
  \bibinfo{author}{Urbakh, M.} \& \bibinfo{author}{Hod, O.}
\newblock \bibinfo{title}{Nanoserpents: Graphene nanoribbon motion on
  two-dimensional hexagonal materials}.
\newblock \emph{\bibinfo{journal}{Nano letters}} \textbf{\bibinfo{volume}{18}},
  \bibinfo{pages}{6009--6016} (\bibinfo{year}{2018}).

\bibitem{ouyang2019mechanical}
\bibinfo{author}{Ouyang, W.} \emph{et~al.}
\newblock \bibinfo{title}{Mechanical and tribological properties of layered
  materials under high pressure: Assessing the importance of many-body
  dispersion effects}.
\newblock \emph{\bibinfo{journal}{Journal of Chemical Theory and Computation}}
  \textbf{\bibinfo{volume}{16}}, \bibinfo{pages}{666--676}
  (\bibinfo{year}{2019}).

\bibitem{thompson2022lammps}
\bibinfo{author}{Thompson, A.~P.} \emph{et~al.}
\newblock \bibinfo{title}{Lammps-a flexible simulation tool for particle-based
  materials modeling at the atomic, meso, and continuum scales}.
\newblock \emph{\bibinfo{journal}{Computer physics communications}}
  \textbf{\bibinfo{volume}{271}}, \bibinfo{pages}{108171}
  (\bibinfo{year}{2022}).

\bibitem{freitas2016nonequilibrium}
\bibinfo{author}{Freitas, R.}, \bibinfo{author}{Asta, M.} \&
  \bibinfo{author}{De~Koning, M.}
\newblock \bibinfo{title}{Nonequilibrium free-energy calculation of solids
  using lammps}.
\newblock \emph{\bibinfo{journal}{Computational Materials Science}}
  \textbf{\bibinfo{volume}{112}}, \bibinfo{pages}{333--341}
  (\bibinfo{year}{2016}).

\bibitem{schebek2026scalable}
\bibinfo{author}{Schebek, M.}, \bibinfo{author}{No{\'e}, F.} \&
  \bibinfo{author}{Rogal, J.}
\newblock \bibinfo{title}{Scalable boltzmann generators for equilibrium
  sampling of large-scale materials}.
\newblock \emph{\bibinfo{journal}{Nature Communications}}
  (\bibinfo{year}{2026}).

\bibitem{sun2015libcint}
\bibinfo{author}{Sun, Q.}
\newblock \bibinfo{title}{Libcint: An efficient general integral library for g
  aussian basis functions}.
\newblock \emph{\bibinfo{journal}{Journal of computational chemistry}}
  \textbf{\bibinfo{volume}{36}}, \bibinfo{pages}{1664--1671}
  (\bibinfo{year}{2015}).

\bibitem{sun2018pyscf}
\bibinfo{author}{Sun, Q.} \emph{et~al.}
\newblock \bibinfo{title}{Pyscf: the python-based simulations of chemistry
  framework}.
\newblock \emph{\bibinfo{journal}{Wiley Interdisciplinary Reviews:
  Computational Molecular Science}} \textbf{\bibinfo{volume}{8}},
  \bibinfo{pages}{e1340} (\bibinfo{year}{2018}).

\bibitem{sun2020recent}
\bibinfo{author}{Sun, Q.} \emph{et~al.}
\newblock \bibinfo{title}{Recent developments in the pyscf program package}.
\newblock \emph{\bibinfo{journal}{The Journal of chemical physics}}
  \textbf{\bibinfo{volume}{153}} (\bibinfo{year}{2020}).

\bibitem{papamakarios2021normalizing}
\bibinfo{author}{Papamakarios, G.}, \bibinfo{author}{Nalisnick, E.},
  \bibinfo{author}{Rezende, D.~J.}, \bibinfo{author}{Mohamed, S.} \&
  \bibinfo{author}{Lakshminarayanan, B.}
\newblock \bibinfo{title}{Normalizing flows for probabilistic modeling and
  inference}.
\newblock \emph{\bibinfo{journal}{Journal of Machine Learning Research}}
  \textbf{\bibinfo{volume}{22}}, \bibinfo{pages}{1--64} (\bibinfo{year}{2021}).

\bibitem{rezende2015variational}
\bibinfo{author}{Rezende, D.} \& \bibinfo{author}{Mohamed, S.}
\newblock \bibinfo{title}{Variational inference with normalizing flows}.
\newblock In \emph{\bibinfo{booktitle}{International conference on machine
  learning}}, \bibinfo{pages}{1530--1538} (\bibinfo{organization}{PMLR},
  \bibinfo{year}{2015}).

\bibitem{durkan2019cubic}
\bibinfo{author}{Durkan, C.}, \bibinfo{author}{Bekasov, A.},
  \bibinfo{author}{Murray, I.} \& \bibinfo{author}{Papamakarios, G.}
\newblock \bibinfo{title}{Cubic-spline flows}.
\newblock \emph{\bibinfo{journal}{arXiv preprint arXiv:1906.02145}}
  (\bibinfo{year}{2019}).

\end{thebibliography}

\end{document}